\documentclass[11pt]{article}
\usepackage{graphicx}
\usepackage{amsmath}
\usepackage{harvard}
\usepackage{a4}

\input{tcilatex}

\begin{document}

\title{Slowly, rotating non-stationary, fluid solutions of Einstein's equations and
their match to Kerr empty space-time}
\author{RJ Wiltshire\thanks{%
Present address: Department of Physics \& Astronomy, Cardiff University,
Queen's Buildings, 5 The Parade, PO Box 913, Cardiff Cf24 3YB. email:
Ron.Wiltshire@astro.cf.ac.uk.} \\
%EndAName
Division of Mathematics and Statistics,\\
The University of Glamorgan, \\
Pontypridd, CF37 1DL. \\
email: rjwiltsh@glam.ac.uk}
\maketitle

\begin{abstract}
A general class of solutions of Einstein's equation for a slowly rotating
fluid source, with supporting internal pressure, is matched using
Lichnerowicz junction conditions, to the Kerr metric up to and including
first order terms in angular speed parameter. It is shown that the match
applies to any previously known non-rotating fluid source made to rotate
slowly for which a zero pressure boundary surface exists. The method is
applied to the dust source of Robertson-Walker and in outline to an interior
solution due to McVittie describing gravitational collapse. The
applicability of the method to additional examples is transparent. The
differential angular velocity of the rotating systems is determined and the
induced rotation of local inertial frame is exhibited.
\end{abstract}

\section{ \ Introduction}

In the period since the discovery of the Kerr [1] metric which describes
analytically, the asymptotically flat, vacuum gravitational field outside a
rotating source in terms of Einstein's field equations, there have been many
attempts to find closed interior solutions which match the exterior
smoothly. In general terms attempts to find solutions have proved
unsuccessful as has been described by Pichon and Lynden-Bell [2]. One
difficulty has been the considerable mathematical complexity in solving
Einstein's equations, see for example, Krasinski [3], Chinea \&
Gonzalez-Romero [4]. This has led to an `embarrassing \ hiatus' according to
Bradley \textit{et al} [5] in the number of \ potential interior solutions
available for matching which \ in turn has contributed to a lack in the
development in the theory of differentially rotating fluid bodies in general
relativity. Even for the case of the remarkable and much quoted, Wahlquist
[6] closed form interior there is no possible fit to the Kerr exterior as
has been recently been shown by Bradley\textit{\ et al }[5]. Only for the
important case \ of thin super-massive rotating discs, supported by internal
pressure have analytic sources for the Kerr metric been found (Pichon and
Lynden-Bell [2]). Yet it is important to develop further the relativistic
theory of rotation since it has considerable potential application in
astrophysics, for example, in the description of the gravitational collapse
of rotating matter, quasars, or potential sources for gravitational
radiation.

To this end one way forward is perhaps to adopt a linearised perturbation
technique wherein, in the first instance, the interior source is rotating
only very slowly. Following the successful match to the Kerr exterior one
would proceed to develop higher order perturbation methods to describe
sources with higher angular velocity. The use of perturbation techniques in
general relativity are of course common. In the context of rotation they
have been applied successfully by Hartle [7] to equilibrium configurations
of cold stars. In the case of non-equilibrium configurations Kegeles [8] has
applied the method to Robertson-Walker dust sources up to the first order in
angular velocity parameter although, the results are somewhat restrictive
and are not suitable for application to sources supported by internal
pressure.

The apppropriate junction conditions for solutions of Einstein's equations
are extensively considered in the literature for example, Misner et al [9],
Mars \& Senovilla [10], Stephani [11], Hernandez-Pastora et al [12] where
the main focus of attention concerns the methods of Darmois [13] and
Lichnerowicz [14] which have been shown to be equivalent by Bonnor and
Vickers [15]. In the Darmois approach it is necessary that the components of
the metric tensor, and also the extrinsic curvature for the Kerr exterior
and the interior source are continuous at the boundary surface. A common \
coordinate description of source and exterior is not \ required. On the
other hand, the Lichnerowicz approach requires a common `admissible'
coordinate system wherein both the metric tensor and its first partial
derivatives are continuous at the boundary surface. Although flexibility and
covariance has ensured the extensive use of the Darmois approach in the
literature, the requirement to define the `admissible' system by
Lichnerowicz has proven to be of benefit for the analysis below. \ 

The main aim here is therefore to develop the perturbation method for
rotating systems up to and including first order terms in the angular
velocity parameter and to find general solutions of Einstein's equations for
perfect fluid bodies which fit the Kerr solution. The solution of Einstein's
equations for the non-rotating system will be assumed given as for example,
the comoving cases presented by McVittie [15], Kustaanheimo [16], Bonnor \&
Faulkes [17], Chakravarty et al [18] and other described by Kramer et al
[19].

Thus in the following the spherically symmetric source will be a given
solution of Einstein's equations described by means of the metric 
\begin{eqnarray}
d\sigma ^{2} &=&e^{2\lambda }d\eta ^{2}-e^{2\mu }d\xi ^{2}-r^{2}d\Omega
^{2}\quad \text{,}  \notag \\
d\Omega ^{2} &=&d\theta ^{2}+\sin ^{2}\theta d\phi ^{2}\quad \text{,}
\label{heath2}
\end{eqnarray}
where $\lambda =\lambda \left( \xi ,\eta \right) $, $\mu =\mu \left( \xi
,\eta \right) $, $r=r\left( \xi ,\eta \right) $ .In the following the
components of this metric will denoted by the tensor $g_{ab}$. The source
boundary surface will be described by the equation: 
\begin{equation}
\xi _{b}=F\left( \eta \right) \quad \text{,}  \label{heath10}
\end{equation}
where $F\left( \eta \right) $ is some function of $\eta $ alone and where
the suffix '$b$' will denote evaluation at the \ boundary throughout.
Lichnerowicz conditions will be used to join the interior to the
Schwarzchild exterior which will be written in the form: 
\begin{eqnarray}
d\sigma ^{2} &=&e^{N}d\Pi ^{2}-e^{-N}d\Sigma ^{2}-\Sigma ^{2}d\bar{\Omega}%
^{2}\quad \text{,}  \notag \\
d\bar{\Omega}^{2} &=&d\bar{\theta}^{2}+\sin ^{2}\bar{\theta}d\bar{\phi}%
^{2}\quad \text{,}  \label{heath14}
\end{eqnarray}
where: 
\begin{equation}
e^{N}=1-\frac{2\bar{m}}{\Sigma }>0\text{ .}  \label{heath16}
\end{equation}
It will be assumed that the coordinate description $\left( \xi ,\theta ,\phi
,\eta \right) $ is a suitable admissible system where the boundary
conditions apply and that this is related to the $\left( \Sigma ,\bar{\theta}%
,\bar{\phi},\Pi \right) $ description of the exterior by means of the
transformation: 
\begin{equation}
\Sigma =\Sigma \left( \xi ,\eta \right) \quad \text{,}\quad \bar{\theta}%
=\theta \quad \text{,}\quad \bar{\phi}=\phi \quad \text{,}\quad \Pi =\Pi
\left( \xi ,\eta \right) \quad \text{.}  \label{heath20}
\end{equation}
It follows that (\ref{heath14}) can be transformed to the form: 
\begin{eqnarray}
d\sigma ^{2} &=&\left( e^{N}\Pi _{\eta }^{2}-e^{-N}\Sigma _{\eta
}^{2}\right) d\eta ^{2}+2\left( e^{N}\Pi _{\xi }\Pi _{\eta }-e^{-N}\Sigma
_{\xi }\Sigma _{\eta }\right) d\xi d\eta  \notag \\
&&-\left( e^{-N}\Sigma _{\xi }^{2}-e^{N}\Pi _{\xi }^{2}\right) d\xi
^{2}-\Sigma ^{2}\left( \xi ,\eta \right) d\Omega ^{2}  \label{heath15}
\end{eqnarray}
where the suffices $\xi $ and $\eta $ mean partial derivatives with respect
to $\xi $ and $\eta $ respectively. The metric components of (\ref{heath15})
will be denoted by $\gamma _{ab}$. In addition units are chosen such that $%
c=1=G$.

The approach adopted here therefore will be at first to apply the
Lichnerowicz junction conditions, in which the components of $g_{ab}$ and,
their first partial derivatives describing a spherically symmetric non
stationary fluid sphere is matched continuously to $\gamma _{ab}$ and, their
first partial derivatives describing the Schwarzchild metric. The conditions
will then be considered in the context of slowly rotating systems. The
results are used to construct new solutions of Einstein's equations which
are applicable for slow rotation and as an example a rotating dust source is
considered and the results of Kegeles [8]. In addition in a further example
the McVittie [15] source is `set' into slow rotation.

\section{Application of the Lichnerowicz junction conditions}

By means of (\ref{heath2}) and (\ref{heath15}) the continuity of the metric
components $g_{kl}$ and $\gamma _{kl}$ , $kl=22$ and first partial
derivatives across the boundary imply that 
\begin{equation}
\Sigma _{b}=r_{b}\quad \text{,}\quad \left\{ \Sigma _{\xi }\right\}
_{b}=\left\{ r_{\xi }\right\} _{b}\quad \text{,}\quad \left\{ \Sigma _{\eta
}\right\} _{b}=\left\{ r_{\eta }\right\} _{b}  \label{heath22}
\end{equation}
where, for any function $X=X\left( \xi ,\eta \right) $: 
\begin{equation}
X_{b}=\left( X\right) _{\xi _{b}=F\left( \eta \right) }\quad \text{,}\quad
\left\{ X_{\xi }\right\} _{b}=\left\{ \frac{\partial X}{\partial \xi }%
\right\} _{\xi _{b}=F\left( \eta \right) }\quad \text{,}\quad \left\{
X_{\eta }\right\} _{b}=\left\{ \frac{\partial X}{\partial \eta }\right\}
_{\xi _{b}=F\left( \eta \right) }  \label{heath28}
\end{equation}
However, for any $X_{b}$ : 
\begin{eqnarray}
\frac{dX_{b}}{d\eta } &=&\left\{ X_{\eta }\right\} _{b}+F_{\eta }\left\{
X_{\xi }\right\} _{b}\quad \text{,}  \notag \\
\left\{ X_{\eta \xi }\right\} _{b} &=&\frac{d\left\{ X_{\xi }\right\} _{b}}{%
d\eta }-F_{\eta }\left\{ X_{\xi \xi }\right\} _{b}  \notag \\
\left\{ X_{\eta \eta }\right\} _{b} &=&\frac{d\left\{ X_{\eta }\right\} _{b}%
}{d\eta }-F_{\eta }\frac{d\left\{ X_{\xi }\right\} _{b}}{d\eta }+F_{\eta
}^{2}\left\{ X_{\xi \xi }\right\} _{b}  \label{heath30}
\end{eqnarray}
and so applying this to $\left\{ \Sigma _{\eta \xi }\right\} _{b}$ and $%
\left\{ \Sigma _{\eta \eta }\right\} _{b}$ it also follows that: 
\begin{eqnarray}
\left\{ \Sigma _{\eta \xi }\right\} _{b} &=&\left\{ r_{\eta \xi }\right\}
_{b}+F_{\eta }\left\{ r_{\xi \xi }\right\} _{b}-F_{\eta }\left\{ \Sigma
_{\xi \xi }\right\} _{b}\quad \text{,}  \notag \\
\left\{ \Sigma _{\eta \eta }\right\} _{b} &=&\left\{ r_{\eta \eta }\right\}
_{b}-F_{\eta }^{2}\left\{ r_{\xi \xi }\right\} _{b}+F_{\eta }^{2}\left\{
\Sigma _{\xi \xi }\right\} _{b}\quad \text{.}  \label{heath46}
\end{eqnarray}
Thus equations (\ref{heath46}) may be used to determine $\left\{ \Sigma
_{\eta \xi }\right\} _{b}$ and $\left\{ \Sigma _{\eta \eta }\right\} _{b}$
once $F$ and $\left\{ \Sigma _{\xi \xi }\right\} _{b}$ have been determined.
In particular, when $F_{\eta }=0$ then $\left\{ \Sigma _{\eta \xi }\right\}
_{b}=\left\{ r_{\eta \xi }\right\} _{b}$ and $\left\{ \Sigma _{\eta \eta
}\right\} _{b}=\left\{ r_{\eta \eta }\right\} _{b}$.

Furthermore from (\ref{heath2}) and (\ref{heath15}) continuity of the
component $g_{kl}$ and $\gamma _{kl}$, $kl=44,14,11$ of the metric across
the boundary gives rise to three further relationships that: 
\begin{eqnarray}
\left\{ \Pi _{\eta }\right\} _{b} &=&\left\{ e^{-N}\left( e^{2\lambda
}+e^{-N}\Sigma _{\eta }^{2}\right) \right\} _{b}^{\frac{1}{2}}\quad \text{,}
\notag \\
\left\{ \Pi _{\xi }\right\} _{b} &=&\left\{ e^{-N}\left( e^{-N}\Sigma _{\xi
}^{2}-e^{2\mu }\right) \right\} _{b}^{\frac{1}{2}}\quad \text{.}
\label{heath57}
\end{eqnarray}
and the physical restriction: 
\begin{equation}
\left\{ e^{2\lambda }\Sigma _{\xi }^{2}-e^{2\left( \mu +\lambda \right)
+N}-e^{2\mu }\Sigma _{\eta }^{2}\right\} _{b}=0\quad \text{,}
\label{heath62}
\end{equation}
where: 
\begin{equation}
\left\{ e^{N}\right\} _{b}=1-\frac{2\bar{m}}{\Sigma _{b}}=1-\frac{2\bar{m}}{%
r_{b}}\quad \text{.}  \label{heath55a}
\end{equation}
Equation (\ref{heath62}) is merely the condition: 
\begin{equation}
\left\{ m\left( \xi ,\eta \right) \right\} _{b}=\bar{m}\text{ .}
\label{heath75}
\end{equation}
where the mass function $m\left( \xi ,\eta \right) $ is as usual defined, in
terms of the Riemann tensor through: 
\begin{equation}
m\left( \xi ,\eta \right) =\frac{r}{2}R_{232}^{3}=\frac{r}{2}\left(
1+e^{-2\lambda }r_{\eta }^{2}-e^{-2\mu }r_{\xi }^{2}\right) \text{.}
\label{heath70}
\end{equation}
This condition may \ be used to simplify (\ref{heath57}) with the result
that: 
\begin{equation}
\left\{ \Pi _{\eta }\right\} _{b}=\left\{ e^{-N}e^{\lambda -\mu }\Sigma
_{\xi }\right\} _{b}\quad \text{,}\qquad \left\{ \Pi _{\xi }\right\}
_{b}=\left\{ e^{-N}e^{\mu -\lambda }\Sigma _{\eta }\right\} _{b}\quad \text{.%
}  \label{heath67}
\end{equation}

Consider now the continuity of the partial derivatives of the metric
components $g_{kl}$, and $\gamma _{kl\text{ }}$for $kl=44,11$ across the
boundary surface. It follows immediately that: 
\begin{eqnarray}
\left\{ \Pi _{\eta \eta }\right\} _{b} &=&\left\{ \frac{\partial }{\partial
\eta }\left[ e^{-\frac{N}{2}}\left( e^{2\lambda }+e^{-N}\Sigma _{\eta
}^{2}\right) ^{\frac{1}{2}}\right] \right\} _{b}\quad \text{,}  \notag \\
\left\{ \Pi _{\xi \xi }\right\} _{b} &=&\left\{ \frac{\partial }{\partial
\xi }\left[ e^{-\frac{N}{2}}\left( e^{-N}\Sigma _{\xi }^{2}-e^{2\mu }\right)
^{\frac{1}{2}}\right] \right\} _{b}\quad \text{,}  \label{heath90}
\end{eqnarray}
and:

\begin{equation}
\left\{ \Pi _{\xi \eta }\right\} _{b}=\left\{ \frac{\partial }{\partial \xi }%
\left[ e^{-\frac{N}{2}}\left( e^{2\lambda }+e^{-N}\Sigma _{\eta }^{2}\right)
^{\frac{1}{2}}\right] \right\} _{b}=\left\{ \frac{\partial }{\partial \eta }%
\left[ e^{-\frac{N}{2}}\left( e^{-N}\Sigma _{\xi }^{2}-e^{2\mu }\right) ^{%
\frac{1}{2}}\right] \right\} _{b}\quad \text{.}  \label{heath94}
\end{equation}
Furthermore direct expansion of the consistency relation $\left\{ \partial
\Pi _{\eta }/\partial \xi \right\} _{b}$ $=\left\{ \partial \Pi _{\xi
}/\partial \eta \right\} _{b}$ in (\ref{heath94}) gives rise to: 
\begin{equation}
\left\{ \Sigma _{\eta \xi }-\mu _{\eta }r_{\xi }-\lambda _{\xi }r_{\eta
}\right\} _{b}=0  \label{heath91a}
\end{equation}
and, using (\ref{heath46}) may be expanded in terms of the Einstein tensor
component $G_{1}^{4}$, to give: 
\begin{equation}
\frac{\left\{ re^{2\lambda }G_{1}^{4}\right\} _{b}}{2}+F_{\eta }\left\{
r_{\xi \xi }-\Sigma _{\xi \xi }\right\} _{b}=0\quad \text{.}
\label{heath91c}
\end{equation}

Furthermore using (\ref{heath90}), (\ref{heath94}) and it follows that \ the
continuity of the partial derivatives of the metric components $g_{kl}$, and 
$\gamma _{kl\text{ }}$for $kl=14$ can also be written as: 
\begin{equation}
0=\left\{ \frac{\partial }{\partial x}\left( e^{-2\mu }\Sigma _{\xi
}^{2}-e^{N}-e^{-2\lambda }\Sigma _{\eta }^{2}\right) \right\} _{b}
\label{heath96}
\end{equation}
where $x=\eta $, $\xi $. With the aid of (\ref{heath46}), these conditions
may also be written in terms of the Einstein tensor components $G_{1}^{4}$
and $G_{4}^{4}$ . Thus: 
\begin{eqnarray}
\left\{ r_{\xi \xi }-\Sigma _{\xi \xi }\right\} _{b}\left\{ r_{\xi }e^{-2\mu
}+F_{\eta }r_{\eta }e^{-2\lambda }\right\} _{b} &=&\frac{r_{b}}{2}\left\{
G_{1}^{4}r_{\eta }-G_{4}^{4}r_{\xi }\right\} _{b}\quad \text{,}  \notag \\
F_{\eta }\left\{ G_{1}^{4}r_{\eta }-G_{4}^{4}r_{\xi }\right\} _{b}+\left\{
G_{4}^{1}r_{\xi }-G_{1}^{1}r_{\eta }\right\} _{b} &=&0\quad \text{.}
\label{heath102}
\end{eqnarray}
The second equation in (\ref{heath102}) may also be expressed in terms of
the mass function to give: 
\begin{equation}
\left\{ F_{\eta }m_{\xi }+m_{\eta }\right\} _{b}=\left\{ \frac{dm}{d\eta }%
\right\} _{b}=0\quad \text{.}  \label{heath106}
\end{equation}
The equations (\ref{heath91c}), and (\ref{heath102}) may be substituted one
into the other to obtain the following simplification: 
\begin{equation}
\left\{ G_{4}^{1}-F_{\eta }G_{4}^{4}\right\} _{b}=0\quad \text{,}\qquad
\left\{ G_{1}^{1}-F_{\eta }G_{1}^{4}\right\} _{b}=0  \label{heath108k}
\end{equation}
\begin{equation}
\left\{ re^{2\mu }G_{4}^{4}\right\} _{b}+2\left\{ \Sigma _{\xi \xi }-r_{\xi
\xi }\right\} _{b}=0\text{.}  \label{heath110}
\end{equation}
These equations may be used to determine $F$, $\left\{ \Sigma _{\xi \xi
}\right\} _{b}$ and also express the restriction (\ref{heath106}). \ 

Note that the above results have been obtained without reference to the
nature of the energy momentum tensor $T_{l}^{k}.$ However if it is now
supposed that the source is a perfect fluid then (\ref{heath108k}), (\ref
{heath106}) become: 
\begin{equation}
p_{b}=0\quad \text{,}\qquad \left\{ u^{1}-F_{\eta }u^{4}\right\} _{b}=0
\label{heath115}
\end{equation}
as expected, where $p\left( \xi ,\eta \right) $ is the source pressure and $%
u^{k}$ are the components of the velocity four-vector. Further in a comoving
description where $u^{1}=0$ then $F\left( \eta \right) $ is a constant,
again as expected.

Thus direct application of the boundary conditions specifies conditions or
limitations on the functions $\Sigma \left( \xi ,\eta \right) $ and $\Pi
\left( \xi ,\eta \right) $ but in no way defines them uniquely. However, the
nature of these transformation functions must be established to complete the
definition of the admissible system. It is straightforward to check that the
transformation functions: 
\begin{equation}
\Sigma \left( \xi ,\eta \right) =r\left( \xi ,\eta \right) \left[ 1-\frac{%
\left\{ e^{2\mu }G_{4}^{4}\right\} _{b}\left( \xi -F\right) ^{2}}{4}%
+\sum_{n=3}^{\infty }D_{n}\left( \xi -F\right) ^{n}\right]  \label{heath210}
\end{equation}
and 
\begin{equation}
\Pi \left( \xi ,\eta \right) =\Pi _{b}+\left\{ \Pi _{\xi }\right\}
_{b}\left( \xi -F\right) +\frac{\left\{ \Pi _{\xi \xi }\right\} _{b}\left(
\xi -F\right) ^{2}}{2}+\sum_{n=3}^{\infty }E_{n}\left( \xi -F\right) ^{n}
\label{heath310}
\end{equation}
with 
\begin{equation}
\qquad \Pi _{b}=\int \left( \left\{ \Pi _{\eta }\right\} _{b}+F_{\eta
}\left\{ \Pi _{\xi }\right\} _{b}\right) d\eta \quad \text{,}
\label{heath311k}
\end{equation}
and $D_{n}=D_{n}\left( \eta \right) $, $E_{n}=E_{n}\left( \eta \right) $, $%
n\geq 3$, are consistent with equations (\ref{heath22}), (\ref{heath46}), (%
\ref{heath67}), (\ref{heath90}), (\ref{heath94}) and (\ref{heath110}). The
ambiguity of $\Sigma \left( \xi ,\eta \right) $ and $\Pi \left( \xi ,\eta
\right) $ is clear through the arbritrary definition of $D_{n}$ and $E_{n}$.

\section{Junction conditions for slowly rotating systems}

Consider now a general first order rotating source given by 
\begin{equation}
d\sigma ^{2}=e^{2\lambda }d\eta ^{2}-e^{2\mu }d\xi ^{2}-r^{2}d\Omega
^{2}-2r^{2}\sin ^{2}\left( \theta \right) q\left( Yd\xi d\phi +Xd\phi d\eta
\right)  \label{cf10}
\end{equation}
where $X=X\left( \xi ,\eta \right) $, $Y=Y\left( \xi ,\eta \right) $ and $q$
is a small angular speed parameter whose square terms and higher are
negligible. Note that up to and including this order the source boundary may
still be written as $\xi =\xi _{b}.$ The exterior metric now the Kerr
solution which is transformed to the form:

\begin{eqnarray}
d\sigma ^{2} &=&\left( e^{N}\Pi _{\eta }^{2}-e^{-N}\Sigma _{\eta
}^{2}\right) d\eta ^{2}+2\left( e^{N}\Pi _{\xi }\Pi _{\eta }-e^{-N}\Sigma
_{\xi }\Sigma _{\eta }\right) d\xi d\eta  \notag \\
&&-\left( e^{-N}\Sigma _{\xi }^{2}-e^{N}\Pi _{\xi }^{2}\right) d\xi
^{2}-\Sigma ^{2}\left( \xi ,\eta \right) d\Omega ^{2}  \notag \\
&&-2\sin ^{2}\left( \theta \right) a\left( \frac{2\bar{m}}{\Sigma }\Pi _{\xi
}d\xi d\phi +\frac{2\bar{m}}{\Sigma }\Pi _{\eta }d\phi d\eta \right)
\label{cf10h}
\end{eqnarray}
where, $a$ is the angular speed parameter with negligibly small quadratic
terms.. The additional continuity conditions applied to $X=X\left( \xi ,\eta
\right) $ and $Y=Y\left( \xi ,\eta \right) $ are obtained by comparing (\ref
{cf10}) with (\ref{cf10h}) so that with \ $a=q$: \ 
\begin{equation}
X_{b}=\left\{ \frac{2\bar{m}\Pi _{\eta }}{\Sigma r^{2}}\right\} _{b}\text{,}%
\qquad Y_{b}=\left\{ \frac{2\bar{m}\Pi _{\xi }}{\Sigma r^{2}}\right\}
_{b}\quad \text{,}  \label{cf40}
\end{equation}

Moreover from (\ref{cf40}) the continuity conditions applied to the
derivative of the metric tensor require: 
\begin{equation}
\left\{ X_{\xi }\right\} _{b}=\left\{ \frac{\partial }{\partial \xi }\left( 
\frac{2\bar{m}\Pi _{\eta }}{\Sigma r^{2}}\right) \right\} _{b}\text{,}\qquad
\left\{ X_{\eta }\right\} _{b}=\left\{ \frac{\partial }{\partial \eta }%
\left( \frac{2\bar{m}\Pi _{\eta }}{\Sigma r^{2}}\right) \right\} _{b}\quad 
\text{,}  \label{cf63g}
\end{equation}
and: 
\begin{equation}
\left\{ Y_{\xi }\right\} _{b}=\left\{ \frac{\partial }{\partial \xi }\left( 
\frac{2\bar{m}\Pi _{\xi }}{\Sigma r^{2}}\right) \right\} _{b}\text{,}\qquad
\left\{ Y_{\eta }\right\} _{b}=\left\{ Y_{b}\right\} _{\eta }=\left\{ \frac{%
\partial }{\partial \eta }\left( \frac{2\bar{m}\Pi _{\xi }}{\Sigma r^{2}}%
\right) \right\} _{b}  \label{cf64t}
\end{equation}
and so using (\ref{heath70}): 
\begin{equation}
\left\{ Y_{\eta }-X_{\xi }\right\} _{b}=\left\{ \frac{6\bar{m}e^{\lambda
+\mu }}{r^{4}}\right\} _{b}\text{ .}  \label{cf48}
\end{equation}

\section{Solution of Einstein's equations for slowly rotating systems}

It will \ now be shown that solutions of Einstein's equations satisfying the
junction conditions do exist. Suppose that Einstein's equations for a
perfect fluid source are written in the form: 
\begin{equation}
G_{b}^{a}=-8\pi T_{b}^{a}\quad \text{,}\quad \quad T_{b}^{a}=\left( \rho
+p\right) u^{a}u_{b}-\delta _{b}^{a}p\quad \text{,}  \label{ebo60p}
\end{equation}
where $\rho $, $p$ are the source density and pressure and $u^{a}$ are the
components of the velocity four-vector with the property that $u^{a}u_{a}=1$
and further suppose, for simplicity that comoving fluid spheres satisfying $%
G_{1}^{1}=G_{2}^{2}$ and $G_{4}^{1}=0$ are given for a non-rotating source.
In addition, it is noted that the source density and supporting internal
pressure for the metric (\ref{cf10}) are not affected by the addition of a
rotation speed parameter up to and including order $q$. However, for the
slowly rotating systems it is necessary to ensure that the components of the
velocity four-vector satisfy $u^{1}=u^{2}$. Now it is straightforward to
show that $G_{3}^{2}=0=G_{2}^{3}$ ,$G_{1}^{2}=0=G_{2}^{1}$ identically for
the metric (\ref{cf10}), whilst $G_{3}^{1}=0=G_{1}^{3}$ will be satisfied so
long as: 
\begin{equation}
\left( Y_{\eta \eta }-X_{\xi \eta }\right) =\left( Y_{\eta }-X_{\xi }\right)
\left( \lambda _{\eta }+\mu _{\eta }-\frac{4r_{\eta }}{r}\right) \quad \text{%
.}  \label{ebo73n}
\end{equation}
Thus Einstein's equations are satisfied whenever: 
\begin{equation}
Y_{\eta }=X_{\xi }+\frac{h\left( \xi \right) e^{\lambda +\mu }}{r^{4}}
\label{cf25}
\end{equation}
where $h$ is an arbitrary function of $\xi $. So comparing (\ref{cf48}) with
(\ref{cf25}) it is seen that the boundary conditions are consistent with
Einstein's equations provided that: 
\begin{equation}
h_{b}=6\bar{m}\text{.}  \label{cf51}
\end{equation}
In addition, the angular velocity $L\left( \xi ,\eta \right) $ of the source
is given by: 
\begin{equation}
L\left( \xi ,\eta \right) =\frac{u^{3}}{u^{4}}=-\frac{qe^{\lambda -\mu
}h_{\xi }}{16\pi r^{4}\left( \rho +p\right) }-qX  \label{cf57y}
\end{equation}
and, a particle moving in \ the field of (\ref{cf10}) will have zero angular
momentum whenever $u_{3}=0$, so that the quantity: 
\begin{equation}
\frac{u_{3}}{u_{4}}=\frac{q\sin ^{2}\theta e^{-\lambda -\mu }h_{\xi }}{16\pi
r^{2}\left( \rho +p\right) }  \label{ebo98k}
\end{equation}
will also be zero for such a particle. It follows that the induced angular
velocity $\Omega \left( \xi ,\eta \right) $ of the inertial frame is given
by: 
\begin{equation}
\Omega \left( \xi ,\eta \right) =-qX\quad \text{,}  \label{ebo98s}
\end{equation}
and that with (\ref{heath67}) and (\ref{cf40}): 
\begin{equation}
\Omega _{b}=-q\left\{ \frac{2\bar{m}e^{-N}e^{\lambda -\mu }\Sigma _{\xi }}{%
\Sigma r^{2}}\right\} _{b}\quad \text{.}  \label{cf45t}
\end{equation}
This confirms that the 'frame dragging' effect decreases inversely with the
cube of $r$ as is well known (for example, Schutz [20]).

Clearly it is now necessary to determine those functions $X\left( \xi ,\eta
\right) $ and $Y\left( \xi ,\eta \right) $ satisfying the junction
conditions which also satisfy (\ref{cf25}). These are established by
writing: 
\begin{eqnarray}
X\left( \xi ,\eta \right) &=&X_{b}+\left( \xi -\xi _{b}\right) \left\{
X_{\xi }\right\} _{b}+\Psi \left( \xi ,\eta \right) \quad \text{,}  \notag \\
Y\left( \xi ,\eta \right) &=&Y_{b}+\Phi \left( \xi ,\eta \right) \quad \text{%
,}  \label{cf45r}
\end{eqnarray}
where $\left\{ \Psi _{\xi }\right\} _{b}=0$ and using (\ref{cf48}) and (\ref
{cf51}) in (\ref{cf25}) gives rise to:

\begin{equation}
\Phi \left( \xi ,\eta \right) =\int \left[ \Psi _{\xi }+\frac{h\left( \xi
\right) e^{\lambda +\mu }}{r^{4}}-\left\{ \frac{6me^{\lambda +\mu }}{r^{4}}%
\right\} _{b}\right] d\eta \quad .  \label{rw4}
\end{equation}
Thus (\ref{cf45r}) with (\ref{rw4}) is solution for $X\left( \xi ,\eta
\right) $ in terms of $Y\left( \xi ,\eta \right) $ and satisfying the
necessary boundary conditions (\ref{cf63g}) to (\ref{cf48}).

\section{A slowly rotating dust cloud}

Suppose that a slowly rotating dust cloud is described by a perturbed
Robertson Walker metric of the type: 
\begin{eqnarray}
d\sigma ^{2} &=&d\eta ^{2}-R^{2}\left( \eta \right) \left( \frac{d\xi ^{2}}{%
1-k\xi ^{2}}+\xi ^{2}d\theta ^{2}+\xi ^{2}\sin ^{2}\left( \theta \right)
d\phi ^{2}\right)  \notag \\
&&-2\xi ^{2}\sin ^{2}\left( \theta \right) R^{2}q\left( Yd\xi d\phi +Xd\phi
d\eta \right)  \label{ebo5}
\end{eqnarray}
It is easy to see that $\left\{ m\left( \xi ,\eta \right) \right\} _{b}=\bar{%
m}$ from (\ref{heath75}) and $p_{b}=0$ from (\ref{heath115}) give rise to: 
\begin{equation}
R_{\eta }^{2}=\frac{2\bar{m}}{\xi _{b}^{3}R}-k\quad  \label{ebo25c}
\end{equation}
The boundary relations for $\Sigma \left( \xi ,\eta \right) $ are: 
\begin{eqnarray}
\Sigma _{b} &=&\xi _{b}R\quad \text{,}\qquad \left( \Sigma _{\xi }\right)
_{b}=R\quad \text{,}\qquad \left( \Sigma _{\eta }\right) _{b}=\xi
_{b}R_{\eta }  \notag \\
\left\{ \Sigma _{\xi \eta }\right\} _{b} &=&R_{\eta }\quad \text{,}\qquad
\left\{ \Sigma _{\eta \eta }\right\} _{b}=-\frac{\bar{m}}{\xi _{b}^{2}R^{2}}%
\quad  \notag \\
\left\{ \Sigma _{\xi \xi }\right\} _{b} &=&\frac{3\bar{m}}{\xi
_{b}^{2}(1-k\xi _{b}^{2})}\text{,}\quad \left\{ e^{N}\right\} _{b}=1-\frac{2%
\bar{m}}{R\xi _{b}}\quad \text{,}  \label{ebo28}
\end{eqnarray}
whilst for $\Pi \left( \xi ,\eta \right) $ : 
\begin{eqnarray}
\Pi _{b} &=&(1-k\xi _{b}^{2})^{\frac{1}{2}}\int e^{-N}d\eta \quad \text{,} 
\notag \\
\left\{ \Pi _{\xi }\right\} _{b} &=&\frac{\left\{ e^{-N}\right\} _{b}\xi
_{b}RR_{\eta }}{(1-k\xi _{b}^{2})^{\frac{1}{2}}}\quad \text{,}\qquad \left\{
\Pi _{\eta }\right\} _{b}=\left\{ e^{-N}\right\} _{b}(1-k\xi _{b}^{2})^{%
\frac{1}{2}}\quad \text{,}\qquad  \notag \\
\left\{ \Pi _{\eta \eta }\right\} _{b} &=&\left\{ \frac{\partial }{\partial
\eta }\left[ \frac{e^{-N}(1-k\xi ^{2})^{\frac{1}{2}}\Sigma _{\xi }}{R}\right]
\right\} _{b}\quad \text{,}\quad \left\{ \Pi _{\xi \eta }\right\}
_{b}=\left\{ \frac{\partial }{\partial \eta }\left[ \frac{e^{-N}R\Sigma
_{\eta }}{(1-k\xi ^{2})^{\frac{1}{2}}}\right] \right\} _{b}  \notag \\
\left\{ \Pi _{\xi \xi }\right\} _{b} &=&\left\{ \frac{\partial }{\partial
\xi }\left[ e^{-N}\left( \Sigma _{\xi }^{2}-\frac{e^{N}R^{2}}{1-k\xi ^{2}}%
\right) ^{\frac{1}{2}}\right] \right\} _{b}\quad  \label{ebo26h}
\end{eqnarray}

Furthermore the final solution (\ref{cf45r}) with (\ref{rw4}) may be written
as: 
\begin{eqnarray}
X\left( \xi ,\eta \right) &=&\frac{2\bar{m}(1-k\xi _{0}^{2})^{\frac{1}{2}%
}\left\{ e^{-N}\right\} _{b}}{\xi _{b}^{3}R^{3}}+\left( \xi -\xi _{b}\right)
\left\{ \frac{\partial }{\partial \xi }\left( \frac{2\bar{m}e^{-N}(1-k\xi
^{2})^{\frac{1}{2}}\Sigma _{\xi }}{\Sigma \xi ^{2}R^{3}}\right) \right\} _{b}
\notag \\
&&+\Psi \left( \xi ,\eta \right) \quad \text{,}\qquad  \notag \\
Y\left( \xi ,\eta \right) &=&\frac{2\bar{m}R_{\eta }\left\{ e^{-N}\right\}
_{b}}{\xi _{b}^{2}(1-k\xi _{b}^{2})^{\frac{1}{2}}R^{2}}+\Phi \left( \xi
,\eta \right) \quad \text{,}  \label{ebo23s}
\end{eqnarray}
with: 
\begin{equation}
\Phi \left( \xi ,\eta \right) =\int \Psi _{\xi }d\eta +\left( \frac{h\left(
\xi \right) }{\xi ^{4}\left( 1-k\xi ^{2}\right) ^{\frac{1}{2}}}-\frac{6\bar{m%
}}{\xi _{b}^{4}\left( 1-k\xi _{b}^{2}\right) ^{\frac{1}{2}}}\right) \int 
\frac{d\eta }{R^{3}}  \label{ebo435}
\end{equation}
This is a general representation of a collapsing, slowly rotating,
Robertson-Walker dust cloud and generalises the results given by Kegeles
(1978) who does not give transparent forms $X\left( \xi ,\eta \right) $ and $%
Y\left( \xi ,\eta \right) .$

\section{Example with non-zero internal pressure}

Consider now, in outline a representative of a broad class of solutions in
which the rotating source has non zero supporting pressure. The chosen
source is the McVittie [15] solution although any spherically symmetric,
comoving solution of Einstein's equations with an added rotation term could
have been used. The only restriction on their use is that any solution must
have a boundary surface where the mass function is constant, or,
equivalently the supporting pressure is zero.

In this case consider (\ref{cf10}) with: 
\begin{eqnarray}
e^{2\mu } &=&\frac{S^{2}}{f^{4}}\left( 1+\frac{f\bar{m}}{2\xi S}\right)
^{4}\quad \text{,}\quad \quad \qquad e^{2\lambda }=\frac{\left( 1-\frac{f%
\bar{m}}{2\xi S}\right) ^{2}}{\left( 1+\frac{f\bar{m}}{2\xi S}\right) ^{2}}%
\qquad \text{,}  \notag \\
\quad r\left( \xi ,\eta \right) &=&\frac{\xi S}{f^{2}}\left( 1+\frac{f\bar{m}%
}{2\xi S}\right) ^{2}\qquad \text{,}\quad \quad f\left( \xi \right) =\left(
1-c_{1}\xi ^{2}\right) ^{\frac{1}{2}}  \label{McV5}
\end{eqnarray}
where $S=S\left( \eta \right) $ and $c_{1}$ is constant. The respective
internal density and supporting pressure are given by:: 
\begin{equation}
8\pi \rho =\frac{K^{5}S_{\eta }^{2}+128\xi ^{3}S^{5}\left( f^{2}-1\right) }{%
K^{5}S^{2}}\quad .  \label{dens}
\end{equation}
\begin{equation}
8\pi p=-\frac{2K^{6}SS_{\eta \eta }+12\xi K^{5}SS_{\eta }^{2}-5K^{6}S_{\eta
}^{2}+256\xi ^{4}S^{6}\left( f^{2}-1\right) }{K^{5}S^{2}\left( 4\xi
S-K\right) }\quad \text{,}  \label{pres}
\end{equation}
where $K=2\xi S+\bar{m}f$. Since this metric is described by a comoving
observer then $F_{\eta }=0$ and the boundary surface is given by the
constant $\xi =\xi _{b}.$This surface is defined through the constancy of
the mass function (\ref{heath75}) at the boundary or by differentiating this
with respect to $\eta $ to obtain the zero pressure boundary condition (\ref
{heath115}). Using the equations one may calculate: 
\begin{equation}
S_{\eta }=\left\{ \frac{8\sqrt{2}\left[ \xi ^{3}S^{5}\left( 2\xi
S-f^{2}K\right) \right] ^{\frac{1}{2}}}{K^{3}}\right\} _{b}\quad \text{.}
\label{McV14}
\end{equation}

The transformation functions, $\Sigma _{b}$, $\left\{ \Sigma _{\xi }\right\}
_{b}$, $\left\{ \Sigma _{\eta }\right\} _{b}$,$\left\{ \Sigma _{\xi \eta
}\right\} _{b}$, $\left\{ \Sigma _{\eta \eta }\right\} _{b}$ and $\left\{
\Sigma _{\xi \xi }\right\} _{b\text{ }}$ may be calculated directly using (%
\ref{heath22}), (\ref{heath46}) and (\ref{heath110}) whilst , $\Pi _{b}$, $%
\left\{ \Pi _{\xi }\right\} _{b}$, $\left\{ \Pi _{\eta }\right\} _{b}$, $%
\left\{ \Pi _{\xi \eta }\right\} _{b}$, $\left\{ \Pi _{\eta \eta }\right\}
_{b}$ and $\left\{ \Pi _{\xi \xi }\right\} _{b\text{ }}$may be determined
easily from (\ref{heath67}), (\ref{heath90}) and (\ref{heath94}). Some
simplification of the results will occur through the use of (\ref{heath70}),
(\ref{McV14}). The resulting somewhat lengthy expressions need not be
reproduced here.

To obtain the corresponding rotating solution it is necessary to calculate
each of $X_{b}$, $\left\{ X_{\xi }\right\} _{b}$ ,$\left\{ X_{\eta }\right\}
_{b}$,$Y_{b}$,$\left\{ Y_{\xi }\right\} _{b}$ and $\left\{ Y_{\eta }\right\}
_{b}$ to ensure a smooth match to empty space-time \ of which $X_{b}$ ,$%
\left\{ X_{\xi }\right\} _{b}$ ,$Y_{b}$, and $\left\{ Y_{\xi }\right\} $ are
also required for the explicit determination of the solution (\ref{cf45r})
and (\ref{rw4}). These calculations are straightforward.

Also note that to obtain the final solution it is necessary to evaluate the
integral (\ref{rw4}) for which the integrand contain terms of the type: 
\begin{equation}
\frac{h\left( \xi \right) e^{\lambda +\mu }}{r^{4}}=\frac{h\left( \xi
\right) f^{6}\left( 1-\frac{f\bar{m}}{2\xi S}\right) }{\xi ^{4}S^{3}\left( 1+%
\frac{fm}{2\xi S}\right) ^{7}}\quad  \label{McV4r}
\end{equation}
where , $h_{b}=6\bar{m}$. Clearly the resulting solution will therefore not
have a closed form. None the less the solution for a rotating version of the
McVittie (1933) has been established formally, although it is interesting to
note that a solution so elegant in the absence of rotation has such an
awkward form when slow rotation is included.

\section{Conclusion}

Solutions of Einstein's equations for slowly rotating time varying sources
supported by internal pressure have been presented. It has been shown that
any known collapsing, or expanding, fluid source known by a comoving
observer not to be rotating may be 'made' to rotate slowly and also matched
smoothly to the Kerr exterior at all times provided that a zero pressure
boundary surface exists. In each case the source rotates with an angular
velocity which is inversely proportional to the sum of its internal density
and pressure and the induced rotation of the inertial frame of reference is
explicitly clear.

It has been shown that rotating solution may be expressed in terms of the
expression $\int e^{\lambda +\mu }r^{-4}d\eta $ and thus it follows that
this must be integrable in closed form for the existence of analytic
solutions. It is interesting to speculate that solutions of Einstein's
equations have previously been found without the need to give the nature of
this expression any consideration. For example it has previously been common
to express comoving systems in terms of the isotropic coordinate $r=\xi
e^{\mu }$ which may not necessarily be an appropriate choice for the
development of rotating descriptions. This matter is currently under
investigation.

The analysis \ here has been presented here with accuracy up to to an
including first order terms in the angular speed parameter. The natural
extension of the work to the matching of second order case with Kerr
space-time, where the source boundary is more complex in nature, will be
reported in the near future. Extensions of this approach to higher orders
may result in an analytic description of sources supported by internal
pressure and which are surrounded by empty space-time.

Finally, although the focus here has been upon slowly rotating compact
bodies it is interesting to note that the Robertson-Walker metric with the
added rotation term (\ref{ebo5}) may be considered as a cosmological model.
In this case the junctions conditions need not be applied. The expressions
for pressure and density remain homogeneous and the angular velocity term (%
\ref{cf57y}) may also be made independent of $\xi $ by choosing: 
\begin{equation}
h\left( \xi \right) =\int \frac{\xi ^{4}d\xi }{\left( 1-k\xi ^{2}\right) ^{%
\frac{1}{2}}}  \label{en1}
\end{equation}
and $X=X\left( \eta \right) .$ The solution of Einstein's equations (\ref
{cf25}) then may be written as: 
\begin{equation}
Y_{\eta }=\frac{h\left( \xi \right) }{\xi ^{4}\left( 1-k\xi ^{2}\right) ^{%
\frac{1}{2}}R^{3}}\text{ .}  \label{en2}
\end{equation}

\section{Acknowledgments}

I would like to record my sincere thanks for the helpful encouragement of
Professor Bill Bonnor at Queen Mary \& Westfield College, London during the
past year and also, Professors Leonid Grishchuk, Mike Edmunds and Peter
Blood for making me so welcome in the Physics and Astronomy Department in
Cardiff. I am also much indebted to those at the University of Glamorgan who
made my visit to Cardiff possible.

\end{document}